\documentclass[aps,prd,nofootinbib,twocolumn,superscriptaddress,preprintnumbers,balancelastpage,longbibliography]{revtex4-1}

\usepackage{amsmath,amssymb,mathtools,bm}
\usepackage{graphicx, color, hepunits}
\usepackage[dvipsnames]{xcolor}
\usepackage{float}
\usepackage{filecontents}
\usepackage{multirow}
 \usepackage{hyperref} 
\hypersetup{
    colorlinks=true,       
    linkcolor=blue,        
    citecolor=blue,        
    filecolor=magenta,     
    urlcolor=blue          
}
\usepackage[utf8]{inputenc}
\usepackage[english]{babel}

\newcommand{\es}[2] {\begin{equation} \label{#1} \begin{split} #2 \end{split} \end{equation}}

\begin{document}

\title{Systematics in the XENON1T data: the 15-keV anti-axion}

\author{Christopher Dessert}
\affiliation{Leinweber Center for Theoretical Physics, Department of Physics, University of Michigan, Ann Arbor, MI 48109 U.S.A.}

\author{Joshua W. Foster}
\affiliation{Leinweber Center for Theoretical Physics, Department of Physics, University of Michigan, Ann Arbor, MI 48109 U.S.A.}

\author{Yonatan~Kahn}
\affiliation{University of Illinois at Urbana-Champaign, Urbana, IL 61801, U.S.A.}

\author{Benjamin R. Safdi}
\affiliation{Leinweber Center for Theoretical Physics, Department of Physics, University of Michigan, Ann Arbor, MI 48109 U.S.A.}

\date{\today}

\begin{abstract}
The XENON1T collaboration~\cite{Aprile:2020tmw} has found an excess of electron recoil events in their Science Run 1 data below $\sim$7 keV with a spectral shape consistent with that expected from a solar-axion-induced signal.  The claimed statistical significance of the solar-axion model over the null hypothesis is 3.5$\sigma$.  In this work we provide evidence for mismodeling in the electron recoil data that 
may decrease the local significance of the axion model to as low as $p \approx 0.1$.  To reach this conclusion, we search for a signal with the spectral template of the solar axion model, but shifted to higher (unphysical) energies above $\sim$7 keV.  We find that the distribution of significances found from this side-band analysis does not follow the expected chi-square distribution, which allows us to quantify the extent to which mismodeling may be affecting the interpretation of the data at energies below $\sim$7 keV.  For example, we find that there is an even higher-significance fit to the data when the solar axion model is shifted upwards in energy by $\sim$15 keV and allowed to have a negative normalization.   
\end{abstract}
\maketitle

\noindent
The XENON1T collaboration has recently reported an excess of electron recoil events in their Science Run 1 (SR1) data in the energy range $\sim$1 - 7 keV over the background expectation~\cite{Aprile:2020tmw}. These results come from an unprecedented 0.65 tonne-years of low-background exposure. The observed spectrum appears consistent with that expected from solar axions, which the collaboration claims are preferred over the null hypothesis by 3.5$\sigma$. In this scenario, axions are produced within the Sun at $\sim$keV energies and subsequently absorbed by the electrons in the XENON1T experiment~\cite{Redondo:2013wwa,Moriyama:1995bz,vanBibber:1988ge,Alessandria:2012mt,Andriamonje:2009dx,Pospelov:2008jk,Gao:2020wer}.  However, the solar interpretation is in strong tension with stellar cooling constraints on axions and axion-like particles~\cite{Giannotti:2015kwo,Viaux:2013lha,Straniero:2018fbv,Diaz:2019kim,Bertolami:2014wua,DiLuzio:2020jjp}. Alternate explanations of the excess include absorption of bosonic dark matter and Tritium decay~\cite{Aprile:2020tmw}, though both  have lower statistical significance than the axion model.

In this work we question the result that the local significance of the axion model over the null hypothesis is 3.5$\sigma$.  In particular, we examine whether the XENON1T data shows evidence for systematic mismodeling at energies above $\sim$7 keV.\footnote{ Throughout this work we refer to any discrepancy between the energy-binned electron recoil data and the background model as mismodeling, but we make no attempt to determine what the source of this mismodeling may be.} Indeed, we find evidence for mismodeling; accounting for this in a data-driven way, the local evidence for the axion model (or, more precisely, the ability to reject the null hypothesis) may drop to the level of $p \approx 0.1$.\footnote{Ref.~\cite{Aprile:2020tmw} also notes a reduction in the significance of the axion signal when an unconstrained Tritium component is included in the background model.}
We frame this discussion in the context of the solar axion spectral template (even though solar axions likely cannot explain the excess~\cite{DiLuzio:2020jjp}, though see~\cite{Gao:2020wer,dent2020inverse}).  Our results should also apply to other explanations of the excess, such as dark matter explanations, which are additionally subject to the look-elsewhere effect.\footnote{We note that quantifying the look-elsewhere effect is not the objective of this paper.}

We emphasize that our analysis is purely statistical in nature: we are completely agnostic as to the physical explanations of the differences from the background modeling of~\cite{Aprile:2020tmw} and the binned electron recoil data. 
The discrepancies pointed out in this work could arise from, for example, the reconstruction of the electron recoil energies from the prompt scintillation (S1) and delayed electroluminesence (S2) data (see, {\it e.g.},~\cite{Aprile:2019dme, Bloch:2020uzh}) or from additional background components that are not properly included in the background model.  

We caution that the analyses presented in this work are simplified relative to the analysis in~\cite{Aprile:2020tmw}.  This is primarily because the full data needed to reproduce the results in~\cite{Aprile:2020tmw} is not readily available.  For example, Ref.~\cite{Aprile:2020tmw} used an un-binned likelihood, but we use a binned likelihood because only binned data for calibrated recoil energies above $\sim$9 keV\footnote{In App.~\ref{App:un-binned} we perform an un-binned analysis of the data below 9 keV.} may be easily extracted from the figures in~\cite{Aprile:2020tmw}.\footnote{Note that we have digitized the data from the figures in~\cite{Aprile:2020tmw}, which also may introduce some amount of human error.}  In addition, Ref.~\cite{Aprile:2020tmw} performed the profile-likelihood procedure, whereby they profiled over their nuisance parameters describing their background model and efficiencies when searching for new-physics signals.  The background is dominated, across most energies and especially below $\sim$35 keV where we focus, by the $^{214}$Pb $\beta$-decay background.  Other important continuum backgrounds arise from decays of $^{133}$Xe, $^{136}$Xe, and $^{85}$Kr.  Mono-energetic backgrounds dominate over the continuum backgrounds in the energy ranges $\sim30 - 75$ keV and $\sim150 - 180$ keV. Since the backgrounds are determined by fits across the entire  $\sim$210 keV energy range, and since the putative axion-induced signal is only a few keV wide, we expect the nuisance parameters for the continuum background components to change very little in the profile likelihood procedure; we verify this explicitly in App.~\ref{App:syst}.  Thus, throughout most of this work (except App.~\ref{App:syst}) we make the approximation that we may take the best-fit background model, as digitized from Figs.~3 and~4 of~\cite{Aprile:2020tmw}, and consider this as a fixed component when searching for evidence in favor of the signal hypothesis.

The remainder of this work is organized as follows.  First, we perform a systematic analysis of the $0-30$ keV data, where we show that while we broadly reproduce the results from~\cite{Aprile:2020tmw} for the axion signal, there is an even more statistically significant \emph{under-fluctuation} when the axion spectral template is shifted by $\sim$15 keV in energy.  We refer to this under-fluctuation as the 15-keV anti-axion signal.  We then make use of the entire $0 - 210$ keV data to investigate the extent of mismodeling across the full energy range.  All analyses and figures in this work may be reproduced using our public code repository~\cite{Dessert2020}. 

\section{Systematic study of the 1--30 keV data}

\begin{figure}[t!]  
\hspace{0pt}
\vspace{-0.2in}
\begin{center}
\includegraphics[width=0.49\textwidth]{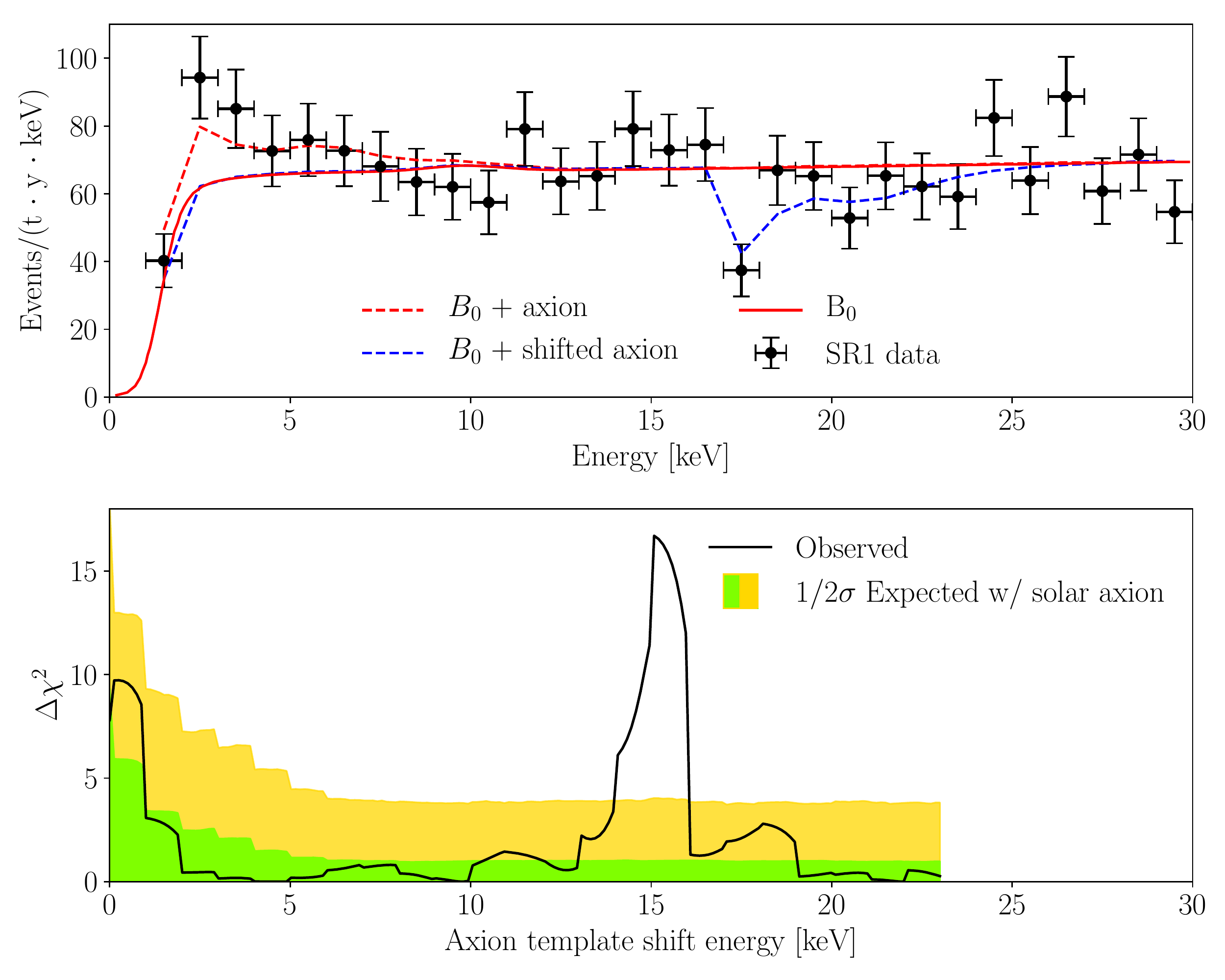}
\caption{ \emph{(Top)} The SR1 data and background model ($B_0$) as digitized from Fig.~4 of~\cite{Aprile:2020tmw}.  We fit the ABC solar axion model, using the detector-level predictions from Fig.~1 of~\cite{Aprile:2020tmw}, to the data to obtain the best-fit signal-plus-background prediction show in dashed red.  However, when we shift the axion signal template to higher energies, we find evidence for an even more significant under-fluctuation at a shift energy of around 15 keV, as indicated by the dashed blue curve. \emph{(Bottom)} The $\Delta \chi^2$ values between the signal and null hypothesis for the analysis where we shift the axion spectral template to higher energies by the amount on the $x$-axis. Note that we cut this test off at $\sim$23 keV to ensure the axion signal may extend over at least 7 keV (the data set presented in~\cite{Aprile:2020tmw} starts at 1 keV and extends to 30 keV). The upper boundary of the green (yellow) band denotes the one-sided 68\% (95\%) containment region for $\Delta \chi^2$ under the hypothesis of an un-shifted axion signal, as constructed from 5000 MC simulations. 
\label{fig:main_0_30}
}
\end{center}
\end{figure}

In the top panel of Fig.~\ref{fig:main_0_30} we reproduce the data and best-fit background model from Fig. 4 of~\cite{Aprile:2020tmw} for the data between 1 and 30 keV.  The deposited energy spectrum for the ABC axion signal, as modeled in~\cite{Redondo:2013wwa}, accounting for the detector energy resolution and efficiency, is found by digitizing the result in Fig. 1 of~\cite{Aprile:2020tmw}.

First, we verify that we may reproduce the evidence for the axion model over the null hypothesis with a simplified analysis framework.  We define the least-squares loss function 
\es{eq:chi2}{
\chi^2(S) = \sum_{i} {\left(B_0^i + S \times S^i - d^i \right)^2 \over \sigma_i^2} \,,
}
where $B_0^i$ is the background model prediction in the $i^{\rm th}$ energy bin, $S^i$ is the ABC axion prediction in that bin (arbitary normalization), $d^i$ is the data (cts/t/yr/keV), and $\sigma_i$ is the statistical uncertainty on the data.  The signal model parameter is $S$, which may take negative values.  We may then compute $\Delta \chi^2 \equiv \chi^2(0) - \chi^2(S_{0})$, where $S_{0}$ is the value that minimizes~\eqref{eq:chi2}.  According to Wilks' theorem, under the null hypothesis we expect $\Delta \chi^2$ to be well-described by the chi-square distribution, with $\sqrt{\Delta \chi^2}$ giving the significance of the signal model over the null hypothesis. We note that the sum in~\eqref{eq:chi2} runs over all energy bins.

Fitting the axion model superimposed on the fixed background model to the data gives the best-fit signal-plus-background model prediction illustrated in dashed red in Fig.~\ref{fig:main_0_30}, with $\Delta \chi^2 \approx 7.8$.  Note that this is of comparable significance to the result quoted in~\cite{Aprile:2020tmw} ($\sim$3.5$\sigma$), but we should not expect the two significances to exactly agree -- Ref.~\cite{Aprile:2020tmw} uses an un-binned likelihood, with nuisance parameters for {\it e.g.} the efficiency, and furthermore Ref.~\cite{Aprile:2020tmw} includes signal contributions from the $^{57}$Fe and Primakoff axion production mechanisms, in addition to the ABC production mechanism considered here.  One indication that the fixed efficiency may be partially responsible for this small discrepancy is that if we remove the first energy bin from the analysis, which is the one most strongly affected by the efficiency, then we find $\Delta \chi^2 \approx 9.7$.  Additionally, as we show in App.~\ref{App:un-binned}, if we perform an un-binned analysis (which is only possible for energies $\lesssim$ 9 keV, as only those events were presented un-binned in Fig. 5 of~\cite{Aprile:2020tmw}) then the significance of the ABC axion signal rises to $\Delta \chi^2 \sim 12$, corresponding to $3.5\sigma$ and consistent with the result in~\cite{Aprile:2020tmw}.

To quantify possible systematic uncertainties from mismodeling,  we repeat the axion search described above for the 1-keV-binned data across the energy ``side-band" at higher energies.  That is, we analyze the same data set but with the axion signal model shifted upwards in energy. Of course, once we shift the energy of the axion signal we are no longer looking for a physically-motivated model.  However, by studying the distribution of significances that we find from this exercise, we can test if the $\Delta \chi^2$ are following the chi-square distribution expected under the null hypothesis. This technique is common in astrophysical searches for dark matter where significances are often determined in a data-driven way, see for example Ref.~\cite{Fermi-LAT:2016uux}.\footnote{In principle this test could be performed with other putative signal templates beyond that of the axion, but as any model which explains the low-energy excess should have a roughly similar spectral shape to that of the ABC axion, we restrict this discussion to that model for simplicity.} Concretely, we shift the axion spectral template by between 0 and 23 keV in steps of 0.25 keV; the upper limit is set by the fact that the data set only extends up to 30 keV, and we want to ensure that the signal model may extend over at least 7 keV.\footnote{Note that we define the shifted signal model at energies at or less than 1 keV above the shift energy to be strictly zero.}

It is worth stressing two points: (i) once we shift the axion model in energy, we are no longer looking for a physical signal, and (ii) even though we are studying the distribution of significances over a wide energy range, our goal in this exercise is primarily to better understand the local significance of the putative axion signal (between 0 and 7 keV) and not to quantify any sort of global significance (along the lines of the look-elsewhere effect).  Note that even though we are not looking for a physical model at higher energies, from a statistical point view the distribution of test statistics associated with a search for this model should still follow the chi-square distribution, if the data is described by the background model to the level of statistical noise. 

The results of the systematic test are illustrated in the bottom panel of Fig.~\ref{fig:main_0_30}, where we show the observed $\Delta \chi^2$ as a function of the energy shift for the axion model.  Interestingly, we observe an even more significant under-fluctuation for a shift energy of around 15 keV, relative to the un-shifted axion search.  The local significance of this (entirely unphysical) ``15 keV anti-axion'' is approximately 4$\sigma$.  The best-fit shifted axion signal is indicated in dashed blue in the top panel of Fig.~\ref{fig:main_0_30}.  This test provides compelling evidence that the background model is not describing the data to the level of statistical noise, for the purposes of the solar axion search, in the 1 to 30 keV energy window, putting aside the putative signal in the $1 - 7$ keV window.  

As a cross-check that the distribution of $\Delta \chi^2$ is expected to be chi-square distributed around 15 keV, even if there is an un-shifted axion signal, we perform 5000 Monte Carlo (MC) simulations under the un-shifted axion hypothesis.  We analyze each of these simulated data sets using the same analysis framework described above, where we shift the axion signal to higher energies.  In the bottom panel of Fig.~\ref{fig:main_0_30} we show the one-sided 68\% and 95\% percentiles of $\Delta \chi^2$ from these simulations.  If the $\Delta \chi^2$ are chi-square distributed and the data is described by the null hypothesis of the background only, then these percentiles should be $\Delta \chi^2 = 1$ and $\Delta \chi^2 = 4$, respectively. We find that this is indeed the case at shift energies above $\sim$7 keV, where the injected un-shifted axion signal drops to zero.  Note that these are one-sided intervals such that {\it e.g.} the interpretation is that in MC, 68\% of the time the $\Delta \chi^2$ value appears below the upper boundary of the green region.  The distribution rises at low energies because the data is simulated with an axion signal, corresponding to the best-fit we find in the real data.  

We note that the significance of the fit with a shift energy of $\sim$15 keV is driven, to a large extent, by the large downward fluctuation in the single bin at $17-18$ keV.  Removing this bin entirely, though, still leaves a shifted-energy downward signal with $\Delta \chi^2 \approx 2$ for a shift of $\sim$15 keV.  This is because all 7 bins from 17 to 23 keV have counts below the background expectation.  It is interesting to compare this with the data in the range $1 - 8$ keV.  In this case the data in the range $2 - 8$ keV is above the background expectation, but it is mostly the $2 - 3$ keV bin that drives the statistical preference for an axion signal; removing the $2 - 3$ keV bin reduces the significance for the axion signal over the background to the level $\Delta \chi^2 \approx 2.7$.  Thus the statistics behind the $1 - 7$ keV excess and the $17 - 23$ keV deficit appear similar, at least the level of the 1-keV binned data.

If we remain agnostic as to the origin of the mismodeling that leads to the 15 keV anti-axion, we can try to estimate the probability that the physical un-shifted axion signal in the $0 - 7$ keV range is also due to mismodeling.  We caution, however, that any argument along these lines should be taken as suggestive only, as we necessarily need to make assumptions that cannot be directly justified by the data.  For example, if we make the assumption that the source of mismodeling that leads to the 15 keV shifted axion signal could also, with equal probability, lead to mismodeling at other energies within the $0 - 30$ keV range, then we may estimate the probability that the un-shifted axion signal arises from mismodeling by noting that $\sim$10\% of the $\Delta \chi^2$ values in Fig.~\ref{fig:main_0_30} for shift-energies above $\sim$7 keV are greater than 9.  That is, while from statistics only we would conclude that $\Delta \chi^2 = 9$ from a two-sided test corresponds to a $p$-value of $p \approx 0.003$, including systematic uncertainties we estimate that $\Delta \chi^2 = 9$ corresponds to a local (two-tailed) $p$-value of $p \approx 0.1$.

In principle, one possible source of systematic mismodeling could be that the background estimate in the range $0 - 30$ keV is too high.  Indeed, if we let the normalization of the background float in the $0 - 30$ keV fit when searching for the un-shifted axion signal, we find that the best-fit background normalization is $\sim$5\% lower than the fiducial normalization.  To test if this shift would resolve the tension at $\sim$15 keV, we construct the profile likelihood for the 15 keV shifted-energy signal in the $0 - 30$ keV energy range while profiling over the normalization of both the background model and the un-shifted axion signal.  That is, in this test the null hypothesis is the model consisting of the background spectral template and the ABC axion template, both with arbitrary normalizations, while the signal hypothesis additionally has a 15-keV shifted axion signal, also with arbitrary normalization.  In this case the $\Delta \chi^2$ value between the null hypothesis and the signal hypothesis is $\sim$14, indicating that there is still more than 3$\sigma$ evidence for a downward fluctuation around 15 keV, even if the normalizations of the background and un-shifted signal models are allowed to float.
We therefore conclude that the putative new physics signal in the $0 - 7$ keV range cannot be taken with the significance claimed in~\cite{Aprile:2020tmw} until the source of the 15 keV under-fluctuation is understood.  

\section{Estimating systematic uncertainties in the 1--210 keV data}

In this section we repeat the analysis described in the previous section, whereby we shift the axion signal to higher energies to search for evidence of background mismodeling, but we extend the analysis to the $1-210$ keV energy range.  The purpose of this section is to investigate whether significant evidence for mismodeling is present in the data at energies other than $\sim$15 keV. 
In this case we make use of the SR1 data presented in Fig.~3 in~\cite{Aprile:2020tmw}, which is unfortunately presented at a lower-resolution energy binning than the Fig.~4 data used in the previous $1-30$ keV section.  In particular, while the $1-30$ keV data is given in 1 keV energy bins, the $0-210$ keV data is presented in energy bins $\sim$2 keV wide.  On the other hand, the Fig. 3 data is presented both for the SR1$_a$ and SR1$_b$ science runs, which allows us to construct a joint likelihood over these data sets.  The SR1$_a$ data set has 55.8 days of exposure, while the SR1$_b$ data has 171.2 days.  The former data set was taken directly after neutron calibrations and is more heavily affected by certain neutron-activated backgrounds.  

We analyze the SR1$_a$ and SR1$_b$ data sets independently for evidence of shifted-energy axion signals, and we also construct a joint likelihood to search for evidence of signals common to both data sets.  To perform the joint likelihood analysis we use $\chi^2(S) \equiv \chi^2_a(S) + \chi^2_b(S)$, where the $\chi^2_{a,b}(S)$ are constructed by modifying~\eqref{eq:chi2} to use the appropriate background model, data counts, and uncertainties.  We assume, as in the previous section, that the background models are fixed, so that we may make use of the best-fit background models given in~\cite{Aprile:2020tmw}.  However, in App.~\ref{App:syst} we show that performing the profile likelihood procedure when searching for the shifted-energy axion signals leads to consistent results.  

\begin{figure}[htb]  
\hspace{0pt}
\vspace{-0.2in}
\begin{center}
\includegraphics[width=0.49\textwidth]{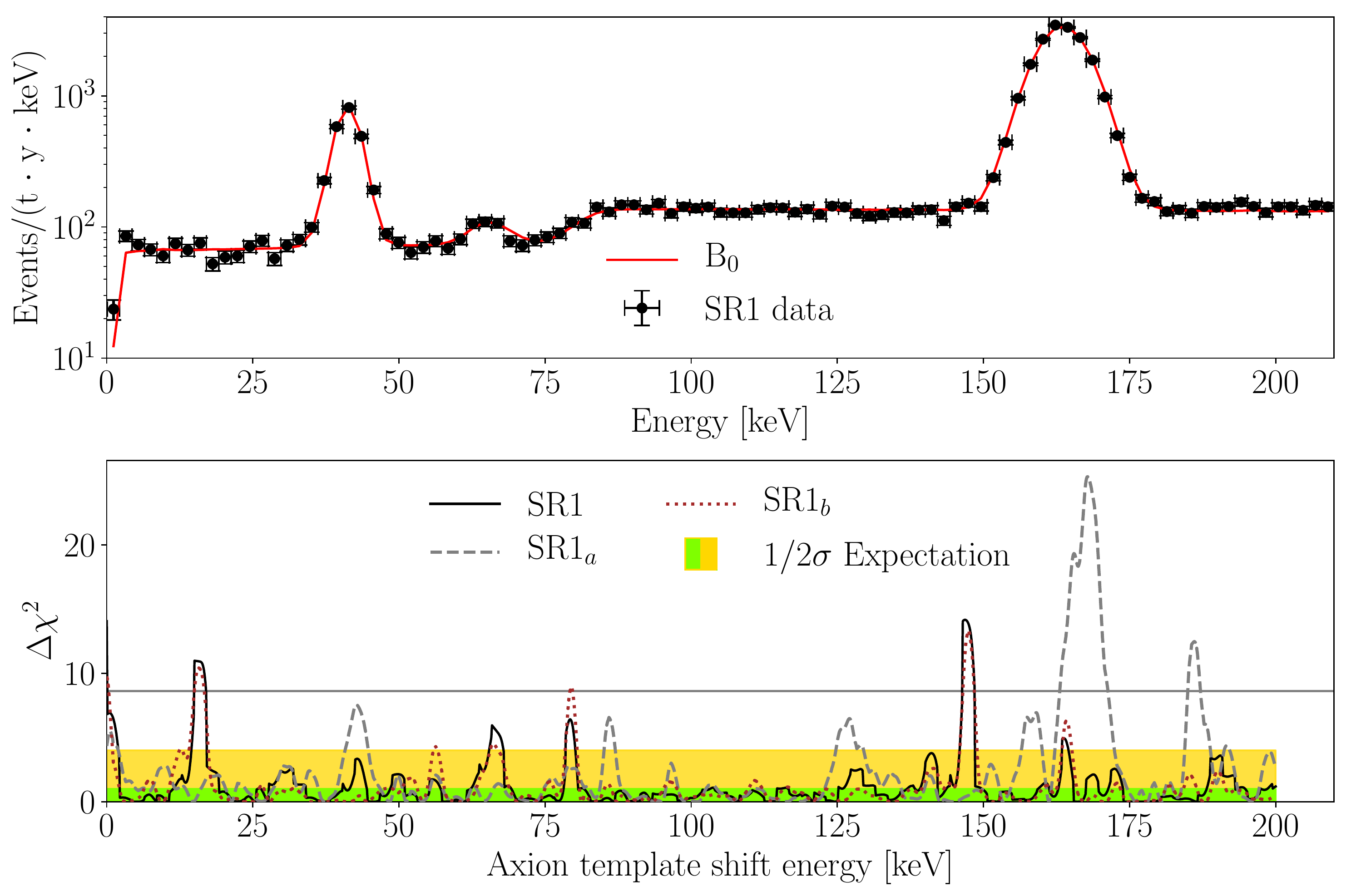}
\caption{As in Fig.~\ref{fig:main_0_30} but using the $1 - 210$ keV SR1$_a$ and SR1$_b$ data as digitized from Fig.~3 in~\cite{Aprile:2020tmw}.   Note that this data has larger bin sizes than the low-energy-only data shown in Fig.~\ref{fig:main_0_30}.  The SR1 $\Delta \chi^2$ curve is constructed from the joint likelihood over the SR1$_a$ and SR1$_b$ data sets.   The most significant evidence for mismodeling in the joint-likelihood analysis is at shift energies $\sim$15 keV and $\sim$148 keV.  The colored bands show the 68\% and 95\% expectations for $\Delta \chi^2$ locally under the null hypothesis, while the horizontal line indicates the maximum global $\Delta \chi^2$ expected at 68\% confidence. }
\label{fig:main_0_200}
\end{center}
\end{figure}

\begin{figure}[htb]  
\hspace{0pt}
\vspace{-0.2in}
\begin{center}
\includegraphics[width=0.45\textwidth]{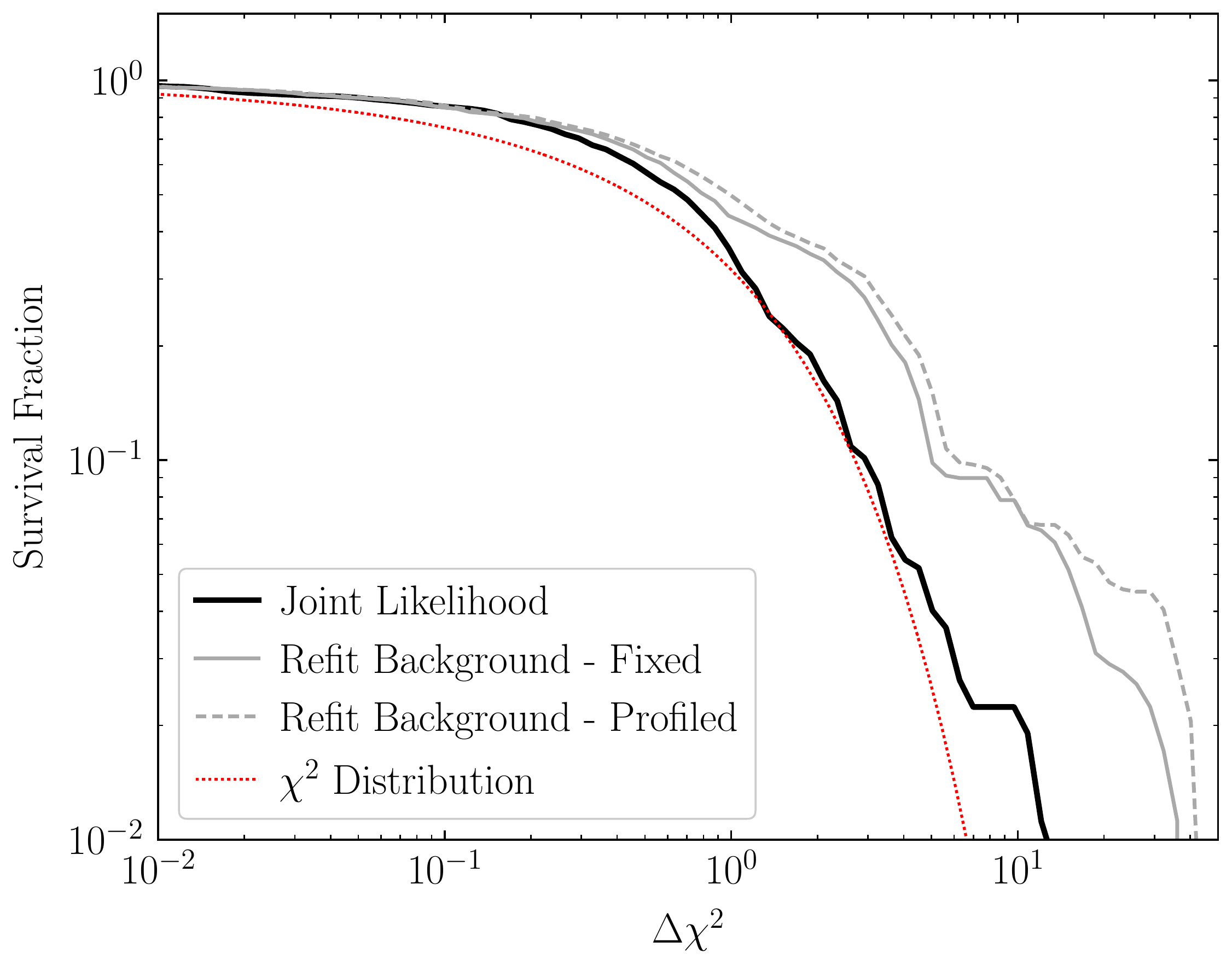}
\caption{The survival fractions for the $\Delta \chi^2$ distributions evaluated with the joint analysis of the SR1$_\mathrm{a}$ and SR1$_\mathrm{b}$ data and with our refit background (see App.~\ref{App:syst}), with and without profiling. In all three cases, the survival fractions for the data lie above the curve for the expected chi-square distribution at high $\Delta \chi^2$ values, indicating an excess of high-significance points compared to statistical expectations alone.}
\label{fig:survival}
\end{center}
\end{figure}

The reproduced data and distribution of significances are illustrated in Fig.~\ref{fig:main_0_200}.  In the top panel we show the combined data set, including both the SR1$_a$ and SR1$_b$ data, along with the combined background model, as reproduced from~\cite{Aprile:2020tmw}.  In the bottom panel we show the distribution of $\Delta \chi^2$ values that we find by shifting the axion spectral template by the indicated value.  Note that we show the values found from analyzing the SR1$_{a,b}$ data sets independently and with the joint likelihood (SR1).  Under the null hypothesis, we find from MC simulations that 68\% of the time the global maximum $\Delta \chi^2$ value across all shifted template energies is below the indicated horizontal line.  In both the joint and individual analyses, however, we observe more high-significance shifted-energy points than expected under the null hypothesis.  This observation is illustrated in Fig.~\ref{fig:survival} (joint likelihood), which shows the fraction of the shifts above 7 keV with $\Delta \chi^2$ larger than the value indicated on the $x$-axis. The survival function lies well above the $\chi^2$ distribution at large $\Delta \chi^2$, indicating an excess of high-significance points. 
For example, in the joint likelihood analysis we observe $\Delta \chi^2 > 9$ (corresponding to more than 3$\sigma$ local significance) for shift energies $\sim$15 keV (as in the previous section) and $\sim$148 keV.  The significance of the $\sim$15 keV shifted fit is smaller in this test than it was in the previous section, but this is likely an artifact of the fact that the putative signal is localized in energy and the 0 - 210 keV data set is presented at lower energy resolution than the $0 - 30$ keV data set.  

Referring to Fig.~\ref{fig:survival} (joint likelihood), we may infer that a value  $\Delta \chi^2 = 9$ has a local significance quantified by the $p$-value $p  \approx 0.02$.  This $p$-value is lower than the estimate in the previous section because the density of high-significance fluctuations is lower in the 1 - 210 keV analysis than in the 1 - 30 keV analysis.  However, the $p$-value 0.02 is significantly higher than that corresponding to a 3$\sigma$ statistical fluctuation (which can be extracted from the chi-square distribution curve in Fig.~\ref{fig:survival}).  Note that in Fig.~\ref{fig:survival} we also show the survival fractions found from two analyses presented in App.~\ref{App:syst}, which produce qualitatively similar results.  In that section we model the combined SR1 data directly (as opposed to using the joint likelihood over the SR1$_a$ and SR1$_b$ data sets), but we construct the background model ourselves by fitting the various contributions to the data over the 1 - 210 keV energy range.  We present results both when the background model is held fixed, which is analogous to the procedure used in our joint likelihood analysis, and when the background is profiled over in the calculation of $\Delta \chi^2$, which is a closer analogue to the procedure in~\cite{Aprile:2020tmw}. 

Analyses of the individual SR1$_a$ and SR1$_b$ data sets themselves each show systematic deviations away from the expectation under the null hypothesis, as indicated in the bottom panel of Fig.~\ref{fig:main_0_200}. However, it is interesting to note that the 15 keV under-fluctuation is more pronounced in the SR1$_b$ data than in the SR1$_a$ data, where it is only around 1$\sigma$ in significance.    

\section{Discussion}

In this work we present evidence of systematic mismodeling in the electron recoil data in~\cite{Aprile:2020tmw}. We come to this conclusion by performing a simplified version of the ABC solar axion analysis, but with the signal spectral template shifted upwards in energy.  Since the real solar axion signal is expected to be confined to approximately $\sim1 - 7$ keV, by shifting the spectral template to higher energies we are able to quantify the extent of mismodeling from the point of view of searching for the physical solar axion signal.  If the background model describes the data to the level of statistical noise, we would expect the distribution of $\Delta \chi^2$ values to follow a chi-square distribution; however, we find significant departures from this distribution in the actual data.  Incorporating these systematic uncertainties in a data-driven fashion likely brings the significance of the true solar-axion signal (in the $1 - 7$ keV energy range) to less than 2$\sigma$ significance.

However, it is difficult to quantify the significance when accounting for mismodeling since the sources of mismodeling -- and thus the chances they affect the $1 - 7$ keV data -- are unknown. In this work we make the implicit assumption that we may use analyses of the data above 7 keV in order to estimate the possibility of mismodeling below 7 keV.  This assumption could break down if, for example, there are reasons to expect the analysis framework to be more reliable below 7 keV than above 7 keV. 

We also stress that while we find that the significance may be less than 2$\sigma$ (and even as low as $p \approx 0.1$) when incorporating systematic uncertainties due to mismodeling, the chance that the $1 - 7$ keV excess is a fluctuation (from either statistics or systematics) is still unlikely.  With this in mind, we are not claiming that no new physics -- either beyond-the-Standard-Model physics or unaccounted-for contaminants like Tritium -- is needed in order to explain the data.  Rather, we are simply suggesting that the significance of the excess may be less than the $3.5\sigma$ claimed in~\cite{Aprile:2020tmw} when the effects of mismodeling are accounted for.  

On the other hand, accounting for mismodeling, the global significance of the $1 - 7$ keV signal is negligible, since there are shift energies that give higher-significance fits to the data (such as for a shift energy of $\sim$15 keV).  The global significance is not the correct significance to use for the solar axion signal, since that signal only physically makes sense in the $1 - 7$ keV energy range and thus the local significance is more appropriate. Explanations of the excess which invoke dark matter of a particular mass, on the other hand, do need to be evaluated with the global significance since they could have in principle given signals at other energies.  Based on our analysis, this implies that there is not sufficient evidence  
at present to justify a dark matter explanation of the excess.  A systematic study along the lines of that presented in this work by the XENON1T collaboration, using the full data set available to them, would help assess the necessity for new physics to explain the low-energy excess.   

\section*{Acknowledgments}
\noindent
{\it    We thank Masha Baryakhtar,  Daniel Baxter, Thomas Boettcher, Johnathon Jordan, Wolfgang Lorenzon, Aaron Pierce, Nicholas Rodd, Peter Sorensen, and Joshua Spitz for helpful comments and feedback. 
 
This  work  was  supported  in  part  by  the  DOE Early  Career  Grant  DESC0019225  and through  computational resources and services provided by Advanced Research  Computing  at the  University  of  Michigan,  Ann Arbor. }

\appendix

\section{Un-binned Likelihood Analysis}
\label{App:un-binned}
\begin{figure}[htb]  
\hspace{0pt}
\vspace{-0.2in}
\begin{center}
\includegraphics[width=0.49\textwidth]{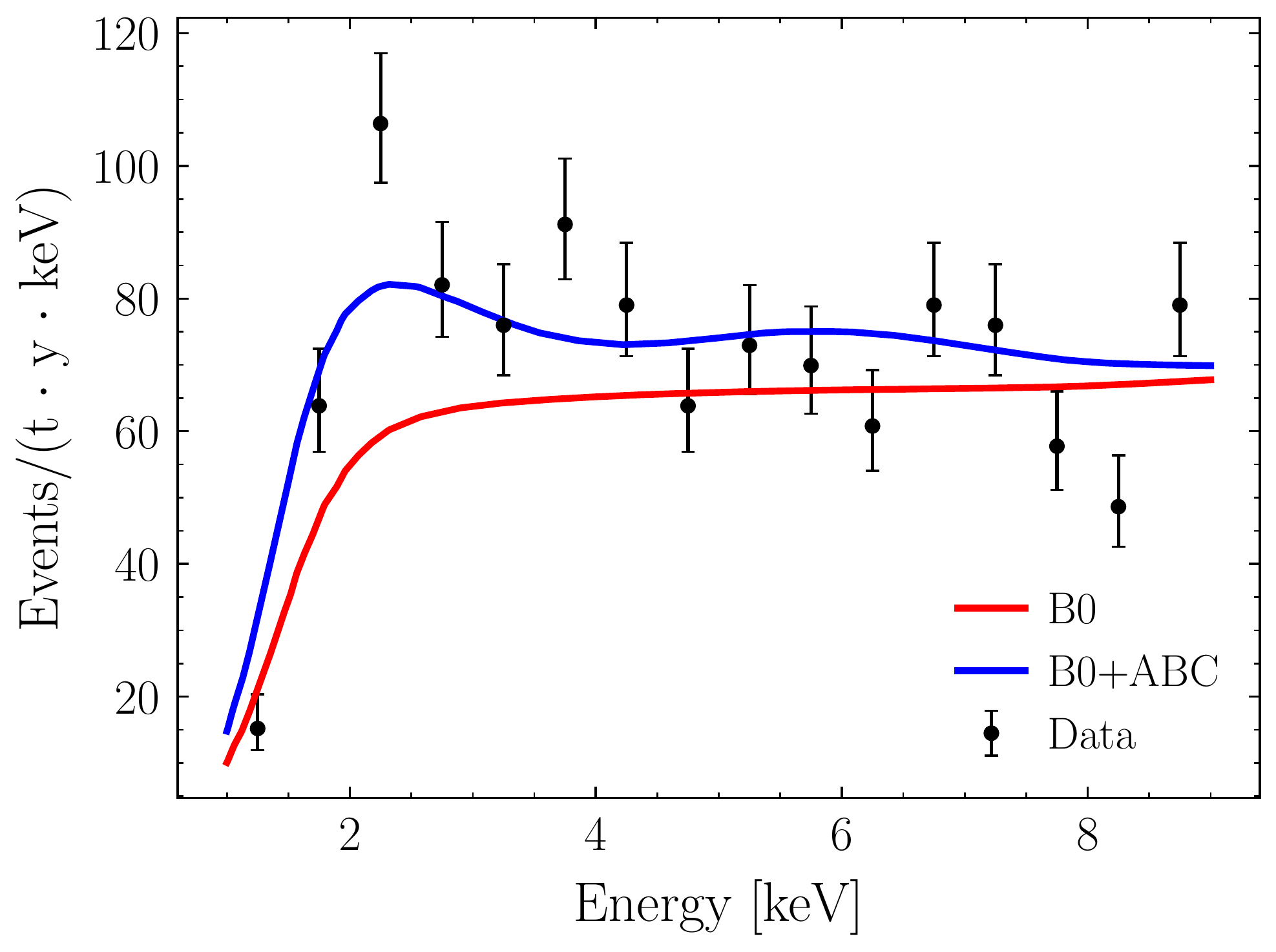}
\caption{The signal model fit to the un-binned events data in the range 1-9 keV. Events are binned at 0.5 keV resolution for the purposes of visualization.}
\label{fig:UnbinnedFit}
\end{center}
\end{figure}

We perform an un-binned analysis using the fixed XENON1T collaboration background model as a test of the detection significance of the un-shifted axion signal. We analyze the data with the un-binned likelihood
\begin{equation}
        \mathcal{L}(\mu_s) = \mathrm{Poiss}(N | \mu_s + \mu_b) \prod_i^N \frac{\mu_b f_b(E_i) + \mu_s f_s(E_i)}{\mu_s + \mu_b}
\end{equation}
over the energies $1-9$ keV, where $\mu_s$ is the free parameter corresponding the number of signal events, $\mu_b$ is the number of background counts fixed by the background model, and $f_b$ and $f_s$ are the fixed probability distribution functions for the energies of events from the background and signal components, respectively. We obtain our best fit for an expected 52 signal counts in the $1-9$ keV range, with fit results shown in Fig.~\ref{fig:UnbinnedFit}. The improvement to the $\chi^2$ statistic for this fit is $\Delta \chi^2 = 12$, which is considerably larger than that which we obtain in the analysis of the binned data, suggesting that an analysis of the un-binned events with the full XENON1T collaboration likelihood may be important for a more detailed characterization of any systematic mismodeling that may occur over the full $0-210$ keV data range. 

\section{Systematic Tests of Analysis Procedure}
\label{App:syst}

\begin{table*}[t]
\begin{tabular}{ |c|c|c|c| } 
 \hline
 Component  & Expected Events & Fitted Events (XENON1T) & Fitted Events (This Work)  \\ \hline
 $^{214}$Pb & (3450, 8530) & $7480 \, \pm 160$ & $ 7496 \pm 243$\\  \hline
 $^{85}$Kr & $890 \pm 150$ & $773 \, \pm 80$ & $902 \pm 144$\\  \hline
 Materials & 323 (fixed) & 323 (fixed) &  323 (fixed) \\ \hline
 $^{136}$ Xe & $ 2120 \pm 210 $ & $ 2150 \, \pm 120 $ & $2121 \pm 20$\\ \hline
 Solar neutrino & $ 220.7 \pm 6.6 $ & $ 220.8 \, \pm 4.7 $ & $220.6 \pm 6.3$\\  \hline
 $^{133}$ Xe & $ 3900 \pm 410 $ & $ 4009 \, \pm 85 $ & $ 4007 \pm 155 $\\ \hline 
 $^{131\mathrm{m}}$ Xe & $ 23760 \pm 640 $ & $ 24270 \, \pm 150 $ & $ 24630 \pm 160 $\\ \hline 
 $^{125}$ I & $ 97.7 \ \pm 33.7 $ & $ 83.0 \, \pm 12.2 $ & $60.4 \pm 24.8$\\ \hline
 $^{83\mathrm{m}}$ Kr & $ 2500 \pm 250 $ & $ 2671 \, \pm 53 $ & $2710 \pm 56$ \\ \hline
 $^{124}$ Xe & $ 157.8 \pm 52.2 $ & $ 149.6 \, \pm 25.1 $ & $ 159.3 \pm 32.2$\\ \hline
\end{tabular}
\caption{A summary of components which contribute to the summed background model used to model the SR1 data over the $0-210$ keV range. Fitting ranges and fit values are presented in terms of the expected number of events contributed by each model components before efficiency correction.}
\label{tab:BackgroundComponents}
\end{table*}

In this Appendix, we perform several tests which validate the methodology used in this work with a primary focus on demonstrating the robustness of the systematic  discrepancies observed under variations on the analysis procedure.

\begin{figure}[htb]  
\hspace{0pt}
\vspace{-0.2in}
\begin{center}
\includegraphics[width=0.495\textwidth]{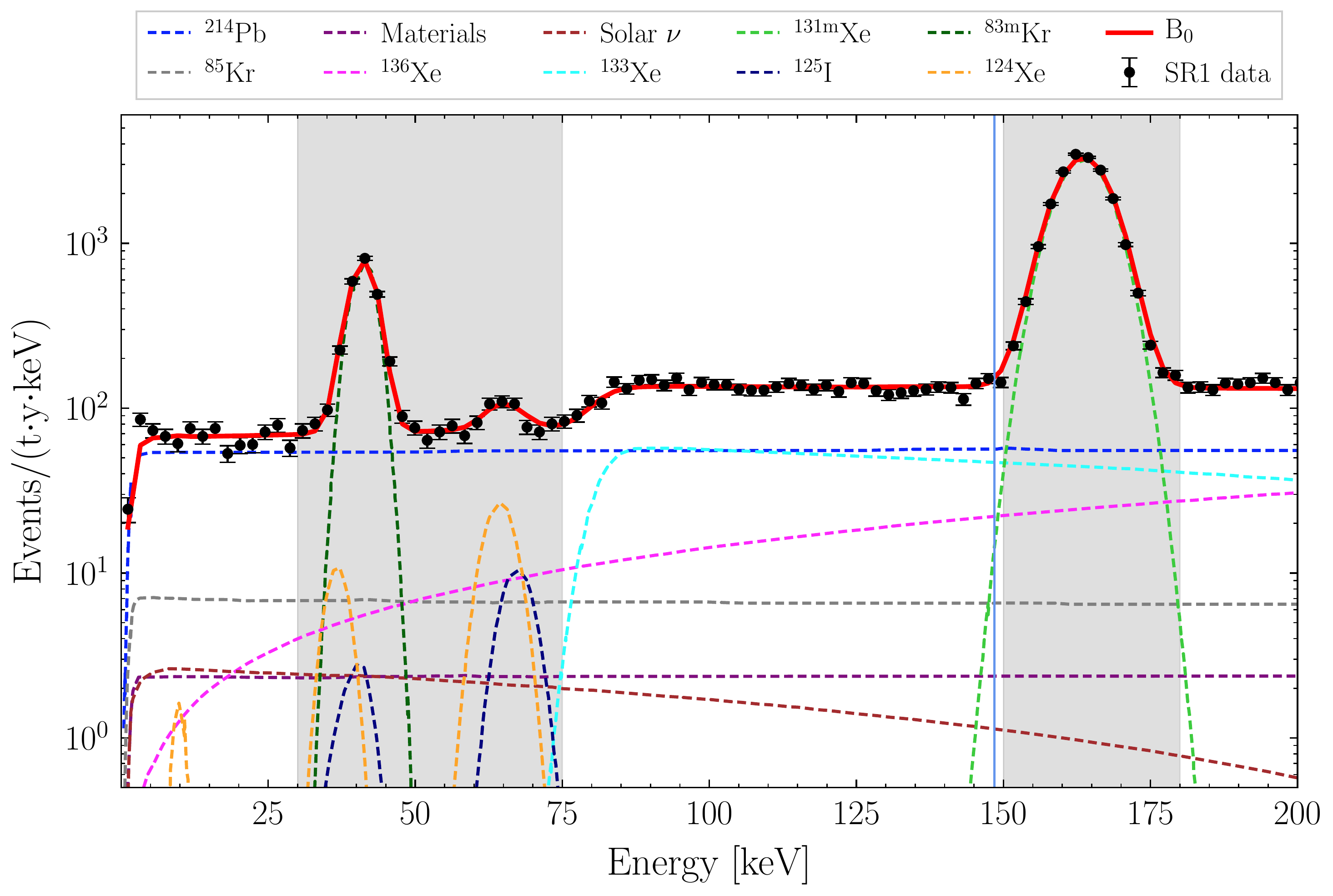}
\caption{Our background model components and summed background model obtained by fitting the model to the SR1 data.}
\label{fig:RefitBackground}
\end{center}
\end{figure}

We begin by refitting each of the background components modeled by the XENON1T collaboration in order to produce our own summed background for the purposes of reanalysis of mismodeling locations.  We perform this analysis on the SR1 data (summed between SR1$_a$ and SR1$_b$) over the $1 - 210$ keV energy range. In Tab.~\ref{tab:BackgroundComponents}, we provide the following for each model component: the priors in counts,\footnote{We use Gaussian priors, except for $^{214}$Pb, which has a linear prior over the quoted range.} the results of the XENON1T collaboration fit to the data, and our own refitted background component amplitudes. We observe qualitative agreement between our own fit and the one obtained by the XENON1T collaboration. Our appropriately normalized individual components and summed background for the SR1 are then depicted in Fig.~\ref{fig:RefitBackground}.  Some discrepancy is expected here since we perform this fit on the summed SR1$_a$ and SR1$_b$ data, while~\cite{Aprile:2020tmw} constructs a joint likelihood between these data sets.  We fit the model directly to the summed data because the priors are not presented separately in~\cite{Aprile:2020tmw} for the SR1$_a$ and SR1$_b$ data sets. 

\begin{figure*}[htb]  
\hspace{0pt}
\vspace{-0.2in}
\begin{center}
\includegraphics[width=0.9\textwidth]{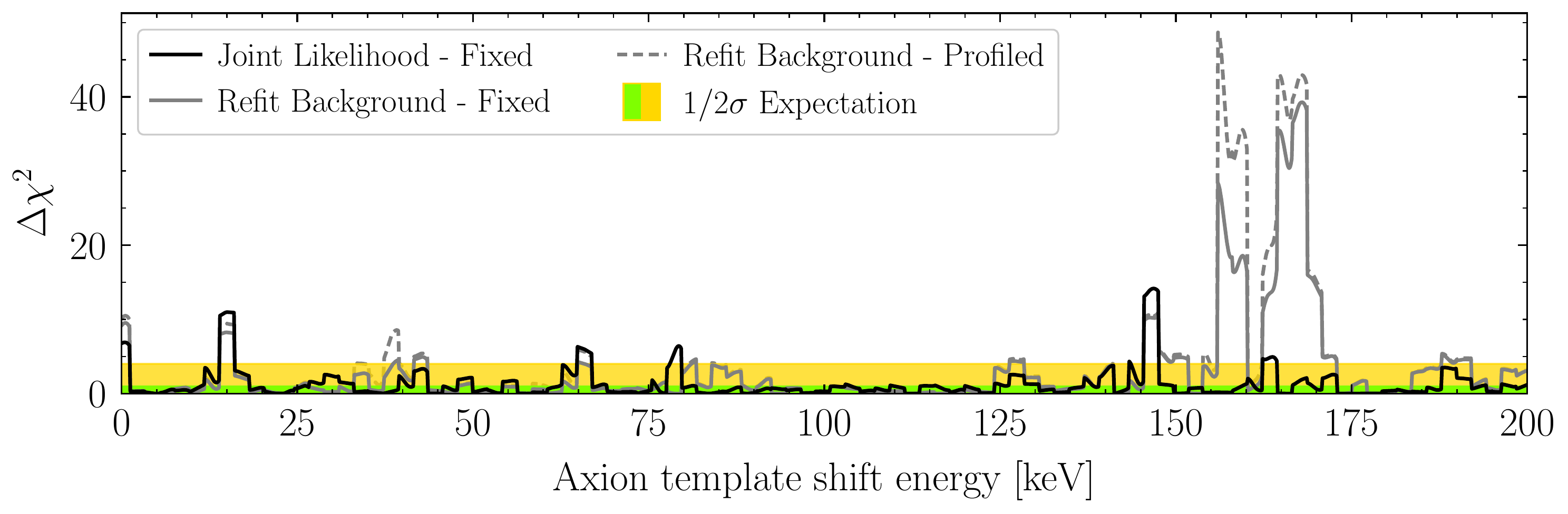}
\caption{As in the bottom panel of  Fig.~\ref{fig:main_0_200}, but with $\Delta \chi^2$ evaluated in the joint likelihood (\textit{black}) and in the stacked data using our refit background with (\textit{dashed grey}) and without (\textit{grey}) profiling of the background components in each fit.}
\label{fig:alt_chi2}
\end{center}
\end{figure*}

After refitting the background, we re-analyze the data over the full energy range with our shifted spectrum analysis for comparison with results obtained with the fixed XENON1T background model. We perform the analysis with and without profiling over the individual background components, with results for $\Delta \chi^2$ as a function of shift energy presented in in Fig.~\ref{fig:alt_chi2}. 

\begin{figure*}[htb]  
\hspace{0pt}
\vspace{-0.2in}
\begin{center}
\includegraphics[width=0.95\textwidth]{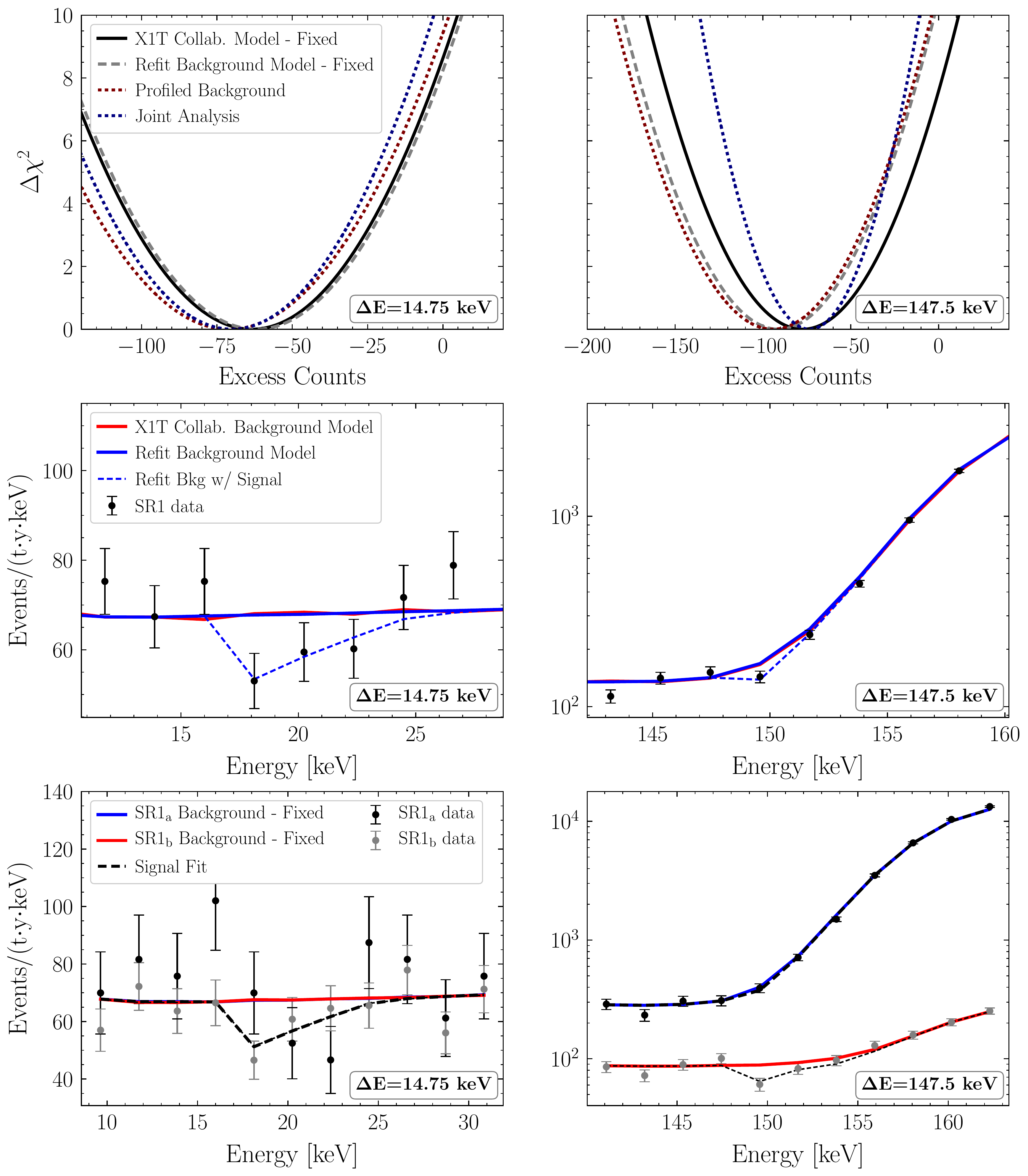}
\caption{Inspection of two identified locations of background mismodeling at approximate energies 15 keV and 150 keV. (\textit{Top}) The $\Delta \chi^2$ associated with the inclusion of a shifted ABC axion spectrum as a function of the expected number of counts contributed by that shifted ABC spectrum. The $\Delta \chi^2$ for the stacked data is evaluated for three choices of background model: the fixed XENON1T collaboration background model, our refit background model fixed at its best fit parameters in absence of a signal model, and with the refit background model fully profiled. We also include the $\Delta \chi^2$ evaluated in the joint analysis of SR1$_\mathrm{a}$ and SR1$_\mathrm{b}$. The energy shift of the ABC spectrum for each analysis is indicated in an inset in the plot. (\textit{Middle}) A comparison of the observed counts with the XENON1T background, our refit background, and our best fit signal under the fixed refit background. (\textit{Bottom}) A comparison of the observed counts, XENON1T background models, and joint signal fit for the SR1$_\mathrm{a}$ and SR1$_\mathrm{b}$ data sets.}
\label{fig:background}
\end{center}
\end{figure*}

The locations of significant mismodelling, which occur at energies of approximately 15 keV and 148 keV, persist after refitting the background. In Fig.~\ref{fig:background} we compare in detail results obtained using our refit backgrounds with those obtained from the fixed XENON1T background models. The high-significance deviations from the background model that we observe are robust with respect to the choice of background model, $i.e.$, our refit background model as compared to the background model from Ref~\cite{Aprile:2020tmw}. We additionally test the effect of performing the profile likelihood procedure when searching for the signal model, with the negligibly impacted results included in Fig.~\ref{fig:background}.

These results justify the simplified analysis used in main results of this work where the background was fixed to the XENON1T collaboration summed background model.  We chose this simplified analysis framework as our fiducial framework because the individual background components were not readily available for all of the data sets analyzed in this work.

\clearpage

\bibliography{axion}

\end{document}